\documentclass[aps,pra,twocolumn,superscriptaddress]{revtex4-2}

\usepackage{bm}
\usepackage{graphicx}
\usepackage{xcolor}
\usepackage{amsmath}
\usepackage{siunitx}

\usepackage{xcolor}

\begin{document}


\title{Fine-structure Raman interferometry with metastable barium \\for magnetically resilient inertial sensing}



\author{Masaki Nagamori}
\affiliation{Department of Physics, Institute of Science Tokyo,\\
2-12-1 Ookayama, Meguro-ku, Tokyo, 152-8550, Japan}

\author{Hiromi Hatano}
\affiliation{Department of Physics, Institute of Science Tokyo,\\
2-12-1 Ookayama, Meguro-ku, Tokyo, 152-8550, Japan}

\author{Toshiyuki Hosoya}
\affiliation{Product Development Center, Japan Aviation Electronics Industry, Ltd.,\\
3-1-1, Musashino, Akishima-shi, Tokyo, 196-8555, Japan}

\author{Yuki Miyazawa}
\affiliation{Institute of Integrated Research, Institute of Science Tokyo,\\
4259 Nagatsuta-cho, Midori-ku, Yokohama, Kanagawa, 226-8501, Japan}

\author{Tomoya Sato}
\affiliation{Institute of Integrated Research, Institute of Science Tokyo,\\
4259 Nagatsuta-cho, Midori-ku, Yokohama, Kanagawa, 226-8501, Japan}

\author{Mikio Kozuma}
\email{kozuma@qnav.iir.isct.ac.jp}
\affiliation{Department of Physics, Institute of Science Tokyo,\\
2-12-1 Ookayama, Meguro-ku, Tokyo, 152-8550, Japan}
\affiliation{Institute of Integrated Research, Institute of Science Tokyo,\\
4259 Nagatsuta-cho, Midori-ku, Yokohama, Kanagawa, 226-8501, Japan}



\date{\today}

\begin{abstract}
Atom-interferometric quantum inertial sensors are promising candidates for inertial navigation systems owing to their high sensitivity and intrinsically low drift. However, their application to next-generation inertial navigation, targeting a drift below \SI{100}{m/month}, requires exceptionally stringent inertial-sensor bias stability. In conventional Raman atom interferometers based on ground-state hyperfine transitions, suppressing inertial-sensor biases arising from quadratic Zeeman shifts requires highly optimized magnetic shielding. In this work, we propose a Raman atom interferometer based on fine-structure transitions between metastable states of barium, which exhibit a quadratic Zeeman sensitivity approximately three orders of magnitude lower than that of conventional alkali-atom Raman transitions. We show that the proposed scheme increases the allowable spatial gradient of the magnetic-field magnitude by a factor of several tens, thereby relaxing the magnetic-field-control requirements from highly optimized shielding and compensation systems to standard magnetic-field-control techniques. 
Furthermore, we experimentally demonstrate Mach--Zehnder interference between metastable fine-structure states of barium, validating a key building block of the proposed inertial-sensing scheme.
\end{abstract}


\maketitle


\section{introduction}

Atom interferometers have been widely used for precision measurements of physical quantities, including determinations of the fine-structure constant through atomic recoil measurements \cite{yu2019,morel2020} and of the Newtonian gravitational constant \cite{fixler2007,lamporesi2008}. 
Owing to their sensitivity to acceleration \cite{Kasevich1991,Kasevich1992,Peters1999} and rotation \cite{gustavson1997,durfeeLongTermStabilityAreaReversible2006}, considerable efforts have been devoted in recent years to the application of atom-interferometric inertial sensors to inertial navigation \cite{Barrett2016, Bongs2019}.
This interest is motivated by the intrinsically high phase sensitivity of matter-wave interferometry, arising mainly from the extremely short de Broglie wavelength of atoms \cite{Cronin2009a}. 
In addition, because atomic properties can serve as absolute references, atom-interferometric sensors are expected to exhibit superior long-term stability compared with conventional inertial sensors.
However, atoms interact with external fields, and the resulting perturbations can limit the measurement accuracy of atom interferometers. 
Among these, magnetic fields constitute one of the dominant sources of systematic error \cite{gauguet2009}. 
Although the linear Zeeman effect can be largely suppressed by employing magnetically insensitive $m=0$ states, residual phase shifts arising from the quadratic Zeeman effect remain a significant challenge \cite{ji2024}.

One approach to mitigating magnetic-field-induced errors is to employ
Bragg diffraction interferometers \cite{muller2008,altin2013}, in which atoms remain in a single internal state throughout the interferometer sequence. Consequently, magnetic-field-induced phase shifts can be substantially reduced.
The magnetic sensitivity can be reduced even further by implementing such interferometers with alkaline-earth or alkaline-earth-like atoms in their $^1S_0$ ground states \cite{mazzoni2015,hosoyaSubrecoilTransverseMomentum2025}, which possess no electronic magnetic moment. 
However, Bragg diffraction requires the transverse velocity spread of the atomic beam to be smaller than the recoil velocity \cite{szigeti2012}. 
As a result, stringent requirements are imposed on the atomic source and optical system, leading to increased experimental complexity \cite{hosoya2023,hosoyaSubrecoilTransverseMomentum2025}.
Atom interferometers based on magnetically insensitive single-photon transitions between the ground and metastable states of alkaline-earth
atoms have also been demonstrated~\cite{Hu2020,Kraft2026}. However, achieving high sensitivity while maintaining the broad
measurement bandwidth required for inertial navigation \cite{tittertonStrapdownInertialNavigation2004}
places stringent requirements on the frequency and phase stability of the laser directly driving the optical transition
\cite{Hu2020,Chiarotti2022,DeRose2023}.

In this work, we propose an alternative approach to mitigating magnetic-field-induced errors by retaining a Raman-based interferometer architecture while employing metastable fine-structure states with intrinsically low quadratic Zeeman sensitivity.
Retaining a Raman-based architecture preserves several important practical advantages.
The Raman beams can be tightly focused to intentionally introduce transit-time broadening, thereby broadening the accepted transverse velocity range, increasing the usable atomic flux, and improving the atom-shot-noise-limited sensitivity.
Another important advantage is that the proposed architecture remains fully compatible with established magnetic-error-suppression techniques, such as $k$-reversal \cite{durfeeLongTermStabilityAreaReversible2006,kwolek2025,satoDigitalClosedloopThermal2026} and composite-light-pulse methods \cite{bergCompositeLightPulseTechniqueHighPrecision2015}.

As a concrete example, we investigate an atom interferometer based on metastable fine-structure states of barium (Ba).
To illustrate the practical significance of the proposed approach, we consider a representative long-term inertial navigation target of \SI{100}{m/month}, corresponding to approximately one order of magnitude higher accuracy than current high-end inertial navigation systems~\cite{paturel2014}. We show that the proposed approach substantially relaxes the magnetic shielding required for achieving this target, shifting it from highly optimized systems \cite{Davis2024BiPlanarCoils,Hobson2022Bespoke} to standard magnetic-field control techniques.

The remainder of this paper is organized as follows. 
In Sec.~\ref{section-theory}, we establish the relationship between interferometer bias and inertial-navigation performance and evaluate the magnetic shielding requirements for alkali-atom interferometers. 
In Sec.~\ref{section-proposal}, we introduce the proposed Ba interferometer and compare its quadratic Zeeman sensitivity and magnetic shielding requirements with those of conventional alkali-atom interferometers. 
In Sec.~\ref{section-experiment}, we experimentally demonstrate
Mach--Zehnder interference between the metastable fine-structure
states of Ba.
Finally, Sec.~\ref{section-conclusion} summarizes the results and discusses future prospects.

\section{Navigation Error Analysis and Magnetic Shielding Requirements}
\label{section-theory}

\subsection{Navigation Error Analysis} 

State-of-the-art strategic-grade inertial navigation systems are approaching unaided navigation performance on the order of one nautical mile per month~\cite{paturel2014}. The next major challenge is to
improve this performance by approximately one order of magnitude, to
$\SI{100}{m}$ per month. Achieving this level of navigation accuracy requires
substantially tighter control of inertial sensor biases. 
To derive the required sensor specifications, we first review the propagation of constant accelerometer and gyroscope biases into navigation errors following the analysis of Groves~\cite{groves2008}. Although quantitative navigation performance should ultimately be evaluated by propagating realistic inertial-sensor error models through the navigation equations along representative trajectories \cite{tittertonStrapdownInertialNavigation2004}, the stationary case isolates the fundamental effects of inertial-sensor biases. 
While particular trajectories may partially cancel or amplify specific error components, the stationary case provides a transparent benchmark for deriving representative sensor requirements.

Constant accelerometer biases produce spurious specific forces that are integrated into the navigation
solution, whereas gyroscope biases generate attitude errors that cause
the gravity vector to be incorrectly projected onto the horizontal
plane.
Since the north and east position errors induced by an accelerometer
bias have identical functional forms, only a representative
horizontal position error is considered in the following discussion.
For a stationary and locally level inertial navigation system, the
representative horizontal position error,
$\delta r_{\mathrm{acc}}$, is given by

\begin{equation}
\delta r_{\mathrm{acc}}(t)
=
\frac{1-\cos(\omega_s t)}
{\omega_s^2}
b_a,
\label{eq:acc_bias}
\end{equation}
where $b_a$ is the constant horizontal accelerometer bias and
$\omega_s$ is the Schuler angular frequency.
The Schuler angular frequency is given by

\begin{equation}
\omega_s
=
\sqrt{\frac{g}{R_E}},
\label{eq:schuler_frequency}
\end{equation}
where $R_E$ is the Earth's radius and $g$ is the gravitational
acceleration. The corresponding Schuler period,
$2\pi/\omega_s$, is approximately \SI{84}{min}.
Equation~(\ref{eq:acc_bias}) therefore shows that the
accelerometer-induced position error remains bounded and oscillates with the Schuler period.
For later discussion, it is convenient to rewrite
Eq.~(\ref{eq:acc_bias}) using 
Eq.~(\ref{eq:schuler_frequency}), yielding

\begin{equation}
\delta r_{\mathrm{acc}}(t)
=
\frac{R_E b_a}{g}
\left[
1-\cos(\omega_s t)
\right].
\label{eq:acc_bias_rewrite}
\end{equation}

For a gyroscope bias, the east position error dominates the long-term
navigation performance because an incorrect estimate of the Earth's
rotation projects the gravity vector onto the horizontal plane. 
Therefore, only the east component is considered in the
following discussion. The corresponding position error $\delta r_{\mathrm{gyro}}$ is

\begin{equation}
\delta r_{\mathrm{gyro}}(t)
=
R_E b_g
\left(
t
-
\frac{\sin(\omega_s t)}
{\omega_s}
\right),
\label{eq:gyro_bias}
\end{equation}
where $b_g$ denotes the corresponding horizontal gyroscope bias.
Unlike Eq.~(\ref{eq:acc_bias_rewrite}), Eq.~(\ref{eq:gyro_bias}) includes a secular term that causes long-term navigation drift.
Since $b_g t$ represents the accumulated attitude error, the factor
$R_E$ converts it into the corresponding east position offset. Thus,
even a minute error in estimating the Earth's rotation rate produces a large eastward drift over long navigation periods.

Using Eqs.~(\ref{eq:acc_bias_rewrite}) and
(\ref{eq:gyro_bias}), the allowable accelerometer and
gyroscope biases required to achieve a position accuracy
of $\Delta r_{\mathrm{err}}$ after a navigation period
$t_m$ are given by
\begin{equation}
b_a
\le
\frac{\Delta r_{\mathrm{err}}g}
{2R_E},
\label{eq:acc_requirement}
\end{equation}
\begin{equation}
b_g
\le
\frac{\Delta r_{\mathrm{err}}}
{R_E t_m},
\label{eq:gyro_requirement}
\end{equation}
respectively. Substituting
$\Delta r_{\mathrm{err}}=\SI{100}{m}$
and
$t_m=\SI{30}{days}$
into Eqs.~(\ref{eq:acc_requirement})
and
(\ref{eq:gyro_requirement})
yields allowable accelerometer and gyroscope biases of
approximately
$\SI{80}{\micro \meter/s^2}$
and
$\SI{1}{\micro deg/h}$,
respectively.
These results establish representative inertial-sensor bias requirements for long-term inertial navigation. In the following section, we derive the corresponding magnetic-field requirements for Raman atom interferometers based on these bias limits.

\subsection{Quadratic Zeeman phase in Raman atom interferometers}

In the previous subsection, we established the allowable accelerometer and gyroscope biases necessary to realize an inertial navigation drift below \SI{100}{m} per month. 
We now derive the magnetic-field-induced phase shift in a Raman atom interferometer and relate it to these bias requirements.
\begin{figure}[htb]
    \centering
    \includegraphics[scale=0.26]{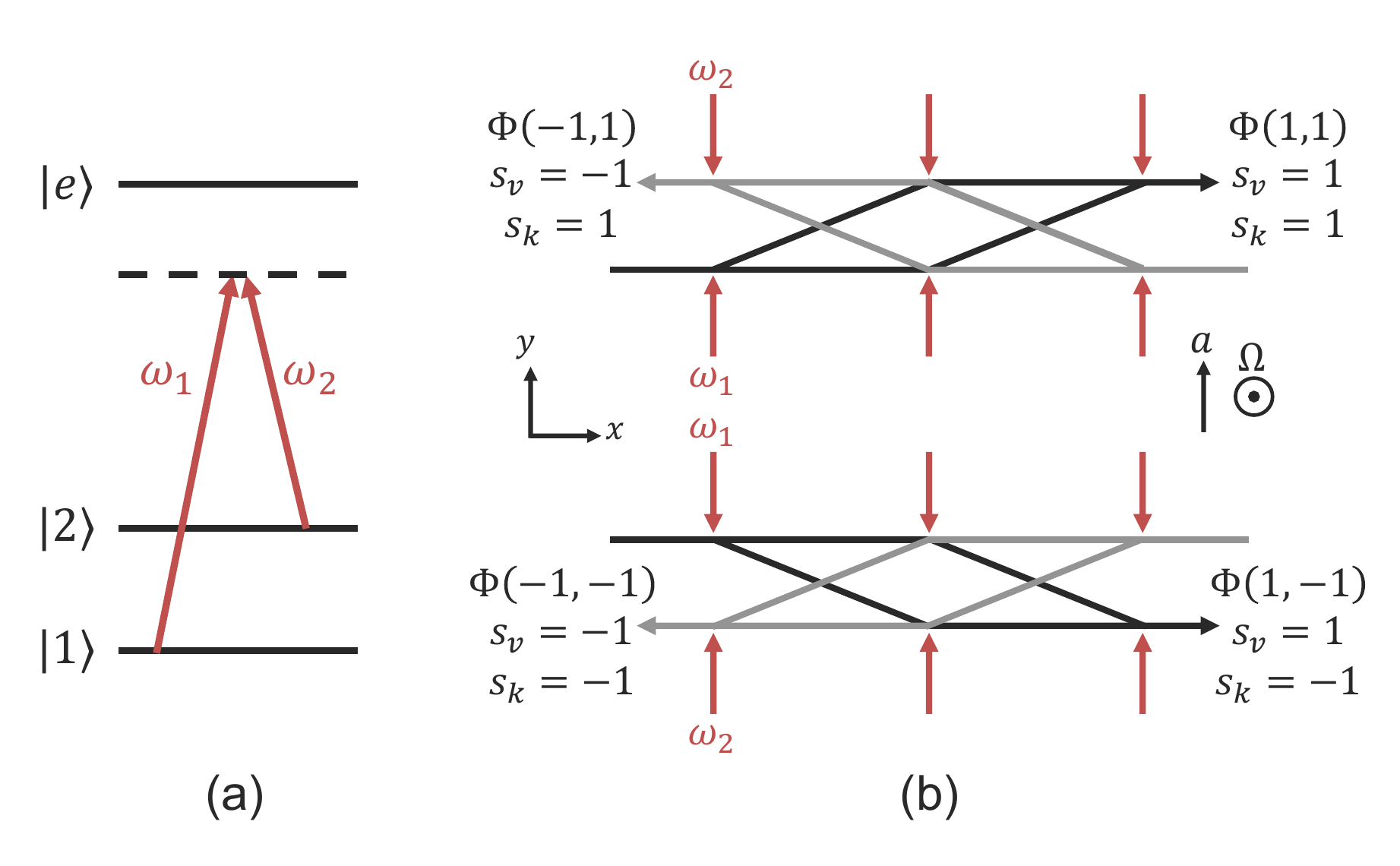}
    \caption{Atom interferometer with counter-propagating beams: (a) Level diagram of stimulated Raman transitions; (b) Raman interferometer configuration employing counter-propagating atomic beams and the $k$-reversal technique for acceleration and rotation measurements.}
    \label{fig_interferometer_Raman_kreverse}
\end{figure}
Specifically, we consider a Raman Mach--Zehnder atom interferometer
employing counter-propagating atomic beams with the
$k$-reversal technique, as illustrated in
Figure~\ref{fig_interferometer_Raman_kreverse}.
The $k$-reversal technique suppresses magnetic phase shifts that are
even under reversal of the effective Raman wave vector, so that only
the residual quadratic Zeeman phase shifts surviving the
$k$-reversal procedure need to be evaluated.
The interferometer is defined in the $xy$ plane. The atomic velocity and effective Raman wave vector are taken to lie along the $x$ and $y$ axes, respectively, and are written as

\begin{equation}
    \bm{v} =s_v v \mathbf{e}_x , \quad \bm{k}_{\mathrm{eff}}=s_k k_{\mathrm{eff}}\mathbf{e}_y,
\end{equation}
where
\begin{equation}
    v=|\bm{v}|,\quad k_{\mathrm{eff}}=|\bm{k}_{\mathrm{eff}}|
\end{equation}
are the magnitudes of the atomic velocity and effective Raman wave vector, respectively, and 
\begin{equation}
    s_v = \pm 1,\quad s_k = \pm 1
\end{equation}
represent the directions of the atomic velocity and effective Raman vector.
The interferometer phase $\Phi(s_v,s_k)$ contains contributions
from various effects, including acceleration, rotation, the Raman laser phase, and magnetic fields. In the following analysis, only the acceleration along the $y$ axis and the rotation about the $z$ axis are considered. For simplicity, these quantities are denoted by $a$ and $\Omega$, respectively.
\begin{equation}
    \Phi(s_v,s_k)
    =
    s_kK_a a
    +
    s_vs_kK_\Omega\Omega
    +
    \Phi_{\mathit{L}}(s_k)
    +
    \Phi_B(s_v,s_k).
\end{equation}
The first two terms correspond to the inertial phases due to
acceleration $a$ and rotation $\Omega$, whose scale factors are given by
\begin{equation}
    K_a=k_{\mathrm{eff}}\left(\frac{L}{v}\right)^2,
    \qquad
    K_\Omega=
    \frac{2k_{\mathrm{eff}}L^2}{v}.
\end{equation}
The remaining terms,
$\Phi_{\mathit{L}}(s_k)$ and
$\Phi_B(s_v,s_k)$,
represent the systematic phase shifts due to the Raman laser phase
and magnetic fields, respectively. 

In the absence of magnetic-field-induced phase shifts, the acceleration
and rotation are obtained from the four ideal interferometer phases
$\Phi^{(0)}(s_v,s_k)$ as
\begin{equation}
    a=
    \frac{1}{4K_a}
    \sum_{s_v=\pm1}
    \left[
        \Phi^{(0)}(s_v,+1)-\Phi^{(0)}(s_v,-1)
    \right],
    \label{eq:acc}
\end{equation}
\begin{equation}
    \Omega=
    \frac{1}{4K_\Omega}
    \sum_{s_v=\pm1}
    s_v
    \left[
        \Phi^{(0)}(s_v,+1)-\Phi^{(0)}(s_v,-1)
    \right].
    \label{eq:gyro}
\end{equation}
While the Raman laser phase
$\Phi_{\mathit{L}}(s_k)$ generally differs between the normal and
$k$-reversed configurations, throughout this paper we assume that
this phase difference is eliminated by an appropriate laser-phase
cancellation technique, such as the digital closed-loop method
described in Ref.~\cite{satoDigitalClosedloopThermal2026}.
Consequently, the phase combinations in Eqs.~\eqref{eq:acc} and \eqref{eq:gyro} are assumed
to be free from laser-phase contributions, so that only
magnetic-field-induced phase shifts need to be considered in the
following analysis.

In the presence of magnetic-field-induced phase shifts, the measured
acceleration and rotation contain systematic biases. These biases are
given by
\begin{equation}
    b_a^{(B)}
    =
    \frac{1}{4K_a}
    \sum_{s_v=\pm1}
    \left[
        \Phi_B(s_v,+1)
        -
        \Phi_B(s_v,-1)
    \right],
\end{equation}
\begin{equation}
    b_g^{(B)}
    =
    \frac{1}{4K_\Omega}
    \sum_{s_v=\pm1}
    s_v
    \left[
        \Phi_B(s_v,+1)
        -
        \Phi_B(s_v,-1)
    \right],
\end{equation}
respectively.
To identify the magnetic-phase components contributing to these biases,
the magnetic phase is decomposed according to its parity under the
reversal of the atomic velocity and the effective Raman wave vector as
\begin{equation}
    \Phi_B(s_v,s_k)
    =
    \Phi_{ee}^{(B)}
    +
    s_v\Phi_{oe}^{(B)}
    +
    s_k\Phi_{eo}^{(B)}
    +
    s_vs_k\Phi_{oo}^{(B)}.
\end{equation}
Here, the first and second subscripts denote the parity under reversal
of the atomic velocity ($s_v$) and the effective Raman wave vector
($s_k$), respectively, with $e$ ($o$) indicating an even (odd)
component.
Since the $k$-reversal operation changes the sign of $s_k$, the
components even under the reversal of the effective Raman wave vector
are canceled. Using the above decomposition, the corresponding accelerometer and
gyroscope biases are given by
\begin{equation}
    b_a^{(B)}
    =
    \frac{\Phi_{eo}^{(B)}}{K_a},
    \label{baB}
\end{equation}
\begin{equation}
    b_g^{(B)}
    =
    \frac{\Phi_{oo}^{(B)}}{K_\Omega},
    \label{bgB}
\end{equation}
Therefore, only the $k$-odd components, $\Phi_{eo}^{(B)}$ and $\Phi_{oo}^{(B)}$, need to be considered in the following analysis.

The quadratic Zeeman shift of each internal state is given by
\begin{equation}
    \Delta \nu_i = \beta_i B^2,
\end{equation}
where $B=|\bm{B}|$ is the magnetic-field magnitude, and
$\beta_1$ and $\beta_2$ are the quadratic Zeeman coefficients
of the internal states $|1\rangle$ and $|2\rangle$, respectively.
The magnetic phase accumulated along the two interferometer arms is therefore expressed as
\begin{equation}
\begin{split}
    \Phi_B &(s_v,s_k) \\
    &=2\pi\int \left[ \beta_U (t) B^2(\bm{r}_U(t)) - \beta_L (t) B^2(\bm{r}_L(t)) \right] dt,
\end{split}
\end{equation}
where $\beta_U(t)$ and $\beta_L(t)$ denote the quadratic Zeeman coefficients corresponding to the internal states occupied by the atoms along the upper and lower interferometer arms, respectively, and $\bm{r}_U(t)$ and $\bm{r}_L(t)$ represent the atomic positions along the corresponding arms.
A bias magnetic field is applied along the effective Raman wave-vector
(the $y$ axis) to define the quantization axis. 
For a magnetically shielded interferometer, we assume that the residual spatial variation of the magnetic-field magnitude is sufficiently small to be approximated by a first-order expansion over the interferometer region,
\begin{equation}
    B(x,y)\simeq B_0+G_xx+G_yy,
\end{equation}
where $B_0$ is the magnetic-field magnitude at the origin. The coefficients $G_x$ and $G_y$ represent the spatial gradients of the magnetic-field magnitude along the atomic-beam direction and the effective Raman wave-vector direction, respectively.
Substituting the atomic trajectories into the above phase integral,
and introducing the differential quadratic Zeeman coefficient
$\beta\equiv\beta_2-\beta_1$,
the remaining magnetic phase components are obtained as
\begin{equation}
\Phi_{eo}^{(B)}
=
\frac{4\pi \beta_2 \hbar k_{\mathrm{eff}} B_0 G_y L^2}
{m v^2},
\end{equation}
and
\begin{equation}
\Phi_{oo}^{(B)}
=
-\frac{2\pi \beta \hbar k_{\mathrm{eff}} G_x G_y L^3}
{m v^2},
\end{equation}
respectively, where \(m\) is the atomic mass.
Using Eqs.~(\ref{baB}) and
(\ref{bgB}), the corresponding acceleration and
gyroscope biases are given by
\begin{equation}
    b_a^{(B)}
    =
    \frac{4\pi \beta_2 \hbar B_0 G_y}
         {m},
\end{equation}
\begin{equation}
    b_g^{(B)}
    =
    -\frac{\pi \beta \hbar G_x G_y L}
          {m v}.
    \label{eq:gyro_bias_mag}
\end{equation}

We consider a realistic transportable atom interferometer with a bias magnetic field of $B_0=\SI{10}{\micro\tesla}$ and an interferometer arm length of $L=\SI{100}{\milli\meter}$. The most probable velocity of the Ba atomic beam used in the experiment described below is approximately $v=\SI{360}{m/s}$ (see Appendix~\ref{app:velocity_distribution}), which is nearly identical to that of an effusive $\mathrm{Rb}$ atomic beam at an oven temperature of \SI{170}{\celsius}. We therefore adopt $v=\SI{360}{m/s}$ for both species to enable a direct comparison under identical interferometer conditions. For $^{87}\mathrm{Rb}$, we consider the ground-state clock transition between $|F=1,m_F=0\rangle$ and $|F=2,m_F=0\rangle$, whose effective quadratic Zeeman coefficient is $\beta=\SI{5.75e10}{Hz/T^2}$ \cite{Wu2014}. Since the two levels exhibit quadratic Zeeman shifts of equal magnitude and opposite sign, we have $\beta_2=\beta/2$.

From the accelerometer bias requirement, the allowable spatial gradient of the magnetic-field magnitude along the $y$ direction is found to be
\begin{equation}
    |G_y|\lesssim \SI{30}{\milli\tesla/m}.
\end{equation}
On the other hand, assuming comparable spatial gradients of the magnetic-field magnitude in the
two directions,
\begin{equation}
    |G_x|=|G_y|=G,
\end{equation}
the gyroscope bias requirement yields
\begin{equation}
    G\lesssim \SI{12}{\micro\tesla/m}.
\end{equation}
Therefore, the required magnetic-shielding performance is determined by
the gyroscope bias requirement.

Maintaining a spatial gradient of the magnetic-field magnitude below approximately $\SI{12}{\micro\tesla/m}$ over long periods is
challenging for a transportable atom interferometer. Residual magnetic
fields and gradients depend on the ambient magnetic environment,
imperfections in the bias-field and shielding systems, and the
orientation-dependent remanent magnetization of the magnetic shields.
Achieving such a level of magnetic-field control often requires
dedicated magnetic-field engineering, including optimized magnetic
shielding and bespoke compensation coils
\cite{Davis2024BiPlanarCoils,Hobson2022Bespoke}. This stringent
requirement motivates the use of atomic species with intrinsically
smaller quadratic Zeeman coefficients. In the next section, we propose
a Raman atom interferometer based on the metastable fine-structure
states of Ba.

\par
\vspace{0.6\baselineskip}

\section{Proposed Barium Atom Interferometer}
\label{section-proposal}
A practical implementation requires an atomic species that combines low quadratic Zeeman sensitivity, readily accessible laser wavelengths, efficient state preparation, and state-selective fluorescence detection.
Although the present concept is not unique to Ba and could, in principle,
be implemented with other atomic species, we selected Ba because it
offers several practical advantages for experimental realization.
In particular, the Raman transition between the metastable states lies in the 
$\SI{1.1}{\micro\meter}$ spectral region and efficient optical pumping into the metastable
states can be achieved without ultraviolet laser light.
These features simplify the optical system and are advantageous for the
future development of compact, robust, and transportable
atom-interferometric inertial sensors.

\begin{figure}[htb]
    \centering
    \includegraphics[scale=0.42]{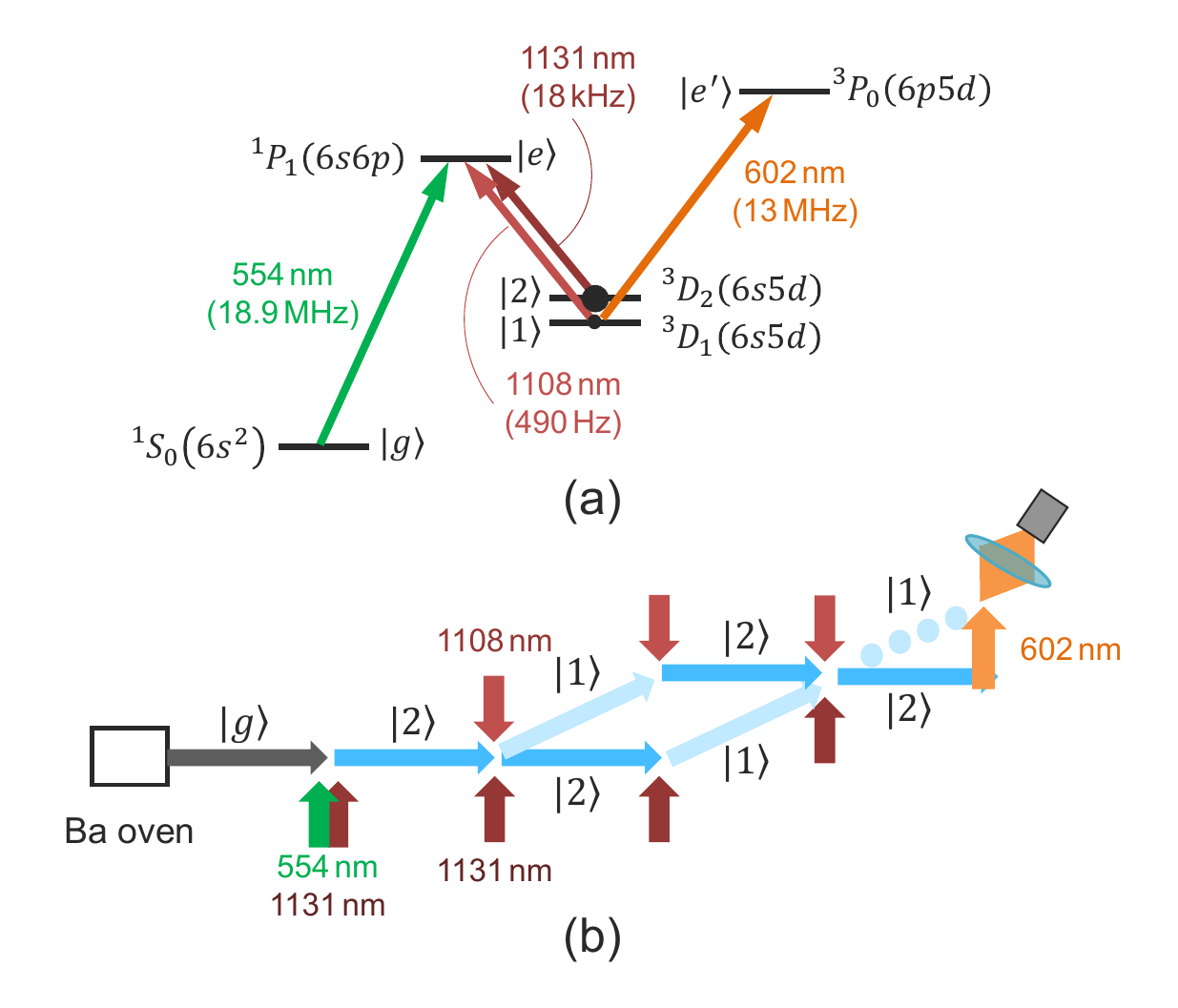}
\caption{
(a) Energy levels and optical transitions of $^{138}$Ba relevant to the proposed atomic-beam interferometer. The labels $|g\rangle$, $|e\rangle$, 
$|e^\prime\rangle$, $|1\rangle$, and  $|2\rangle$
are used throughout the paper.
Values in parentheses give $\Gamma/2\pi$, where $\Gamma$ is the spontaneous decay rate of the corresponding optical transition \cite{Dammalapati2009, Kulaga2001}.
(b) Conceptual schematic of the complete $^{138}$Ba atomic-beam
interferometer, including atomic-beam generation, state preparation,
Raman interferometry, and state-selective detection.
}
    \label{fig_proposal_Ba_interferometer}
\end{figure}

The proposed Ba atom interferometer is based on the fine-structure
levels shown in Fig.~\ref{fig_proposal_Ba_interferometer}(a).
Because bosonic $^{138}$Ba (natural abundance: 
\SI{71.7}{\percent}) has a nuclear spin of $I=0$, it possesses no hyperfine structure. Unlike conventional alkali-atom interferometers, which employ Raman
transitions between ground-state hyperfine levels, the proposed scheme
uses a Raman transition between the fine-structure states
$|1\rangle$ (${}^3D_1\,(6s5d)$) and
$|2\rangle$ (${}^3D_2\,(6s5d)$).
Note that the ground-state hyperfine splitting of $^{87}$Rb is
\SI{6.8}{GHz}, whereas the energy separation between
$|1\rangle$ and $|2\rangle$ in $^{138}$Ba is \SI{5.4}{THz}.
The large fine-structure splitting is expected to substantially suppress
the differential quadratic Zeeman shift and thereby reduce the magnetic
sensitivity of the interferometer.

The differential quadratic Zeeman coefficient of the
$|1,m_J=0\rangle \leftrightarrow |2,m_J=0\rangle$
transition is calculated to be
\SI{3.7e7}{Hz/T^2}
using the $LS$-coupling approximation~\cite{Beverini1998}.
This value is substantially smaller than that for the ground-state
hyperfine Raman transition in $^{87}$Rb,
$\SI{5.75e10}{\hertz/\tesla^2}$.
For comparable atomic-beam velocities, Eq.~\eqref{eq:gyro_bias_mag} shows that the
permissible gradient scales as $\sqrt{m/|\beta|}$. 
The corresponding requirement is therefore relaxed
from $\SI{12}{\micro\tesla/m}$ for $^{87}$Rb to $\SI{570}{\micro\tesla/m}$ for $^{138}$Ba under otherwise
identical operating conditions.
Although the improvement factor is approximately 50, the relaxed requirement moves the allowable gradient from a regime requiring highly optimized magnetic shielding and compensation to one that is readily achievable with standard magnetic-field control techniques.

As illustrated in Fig.~\ref{fig_proposal_Ba_interferometer}(b),
a thermal beam of bosonic $^{138}$Ba atoms is generated from an
effusive oven.
The atoms are first transferred from the ground state
$|g\rangle$ (${}^1S_0\,(6s^2)$)
to the metastable state
$|2\rangle$
by a Raman transition driven by 554- and 1131-nm laser fields through the
intermediate state
$|e\rangle$ (${}^1P_1\,(6s6p)$). 
The metastable states
$|1\rangle$ and $|2\rangle$
are predicted to have radiative lifetimes of approximately
\SI{60}{s} and more than \SI{60}{s}, respectively, according to relativistic
multiconfiguration Dirac--Fock calculations \cite{Migdalek1990}.
The Raman transition is driven by 1108- and 1131-nm laser fields via
the intermediate state $|e\rangle$.
The first Raman pulse acts as a $\pi/2$ pulse, coherently splitting the
incident atomic wave packet while simultaneously transferring the
internal state. A subsequent $\pi$ pulse exchanges the two
interferometer arms, thereby acting as an atomic mirror, and a final
$\pi/2$ pulse recombines the wave packets to close the interferometer.
After the interferometer sequence, laser light at \SI{602}{nm}
resonant with the $|1\rangle$--$|e'\rangle$ transition is applied,
where $|e'\rangle$ denotes the $^3P_0\,(6p5d)$ state.
The resulting fluorescence is proportional to the population in
$|1\rangle$ and provides the interferometer output signal.

Another practical advantage of the proposed scheme is that all
$^{138}$Ba atoms in the ground state are naturally in the
$|g,m_J=0\rangle$ state. Appropriate Raman polarizations then enable
selective transfer to the $|2,m_J=0\rangle$ state. Consequently,
unlike conventional alkali-atom Raman interferometers, no optical
pumping into a magnetically insensitive Zeeman sublevel is required.
Moreover, because the interferometer operates exclusively on metastable
atoms and the fluorescence signal is detected only from the metastable
transition, ground-state atoms do not contribute to the detected
signal. This intrinsic state selectivity substantially suppresses the
detection background compared with conventional alkali-atom thermal-beam
interferometers. In addition, the optical preparation of metastable
atoms can, in principle, be combined with transverse velocity selection,
providing an additional degree of freedom for optimizing the coherence
and signal-to-noise ratio of the atomic beam.

The choice of Ba also offers significant advantages for the laser implementation.
The Raman wavelengths employed in the present scheme, 1108 and \SI{1131}{nm},
both lie in the $\SI{1}{\micro \meter}$ spectral region. Their relative optical phase
can therefore be stabilized by phase-locking both lasers to a common
optical frequency comb operating near $\SI{1}{\micro m}$ \cite{Gunton2015}.
Because of their close wavelengths, the two Raman beams can be combined
into a single-mode optical fiber and delivered through a common optical path.
This configuration is fully compatible with the retroreflected Raman
geometry developed in
Refs.~\cite{kwolek2025,satoDigitalClosedloopThermal2026}.
In this geometry, the Raman beams are directed at a small angle
with respect to the direction perpendicular to the atomic beam and
retroreflected by a mirror. The resulting Doppler shift selectively
brings only one of the two oppositely directed Raman transitions into
resonance, thereby defining a unique effective Raman wavevector.
Accordingly, the same implementation can be directly applied to the
present scheme while maintaining the required relative phase stability.

\par
\vspace{0.6\baselineskip}

\section{PROOF-OF-PRINCIPLE DEMONSTRATION OF MACH--ZEHNDER INTERFERENCE\label{section-experiment}}

Recent advances in neutral-atom quantum computing have enabled coherent Raman manipulation between metastable fine-structure states of Sr for fast qubit control~\cite{Pucher2024,Unnikrishnan2024}. However, the extension of such coherent manipulation to Mach--Zehnder atom interferometry has not yet been demonstrated. Here, we demonstrate coherent Mach--Zehnder interference between metastable fine-structure states of Ba as a first step toward realizing the proposed interferometer.
For this proof-of-principle demonstration, we adopted two conditions that differ from those of the full implementation to facilitate the experiment.
First, the 1108- and 1131-nm Raman lasers were frequency-locked to a common optical frequency comb using delay-line discriminators for their respective beat notes \cite{Hisai2018}, while their relative optical phase was left unstabilized. 
Second, co-propagating Raman beams were employed to reduce sensitivity to environmental vibrations and other inertial perturbations.

\begin{figure}[htb]
    \centering
    \includegraphics[scale=0.32]{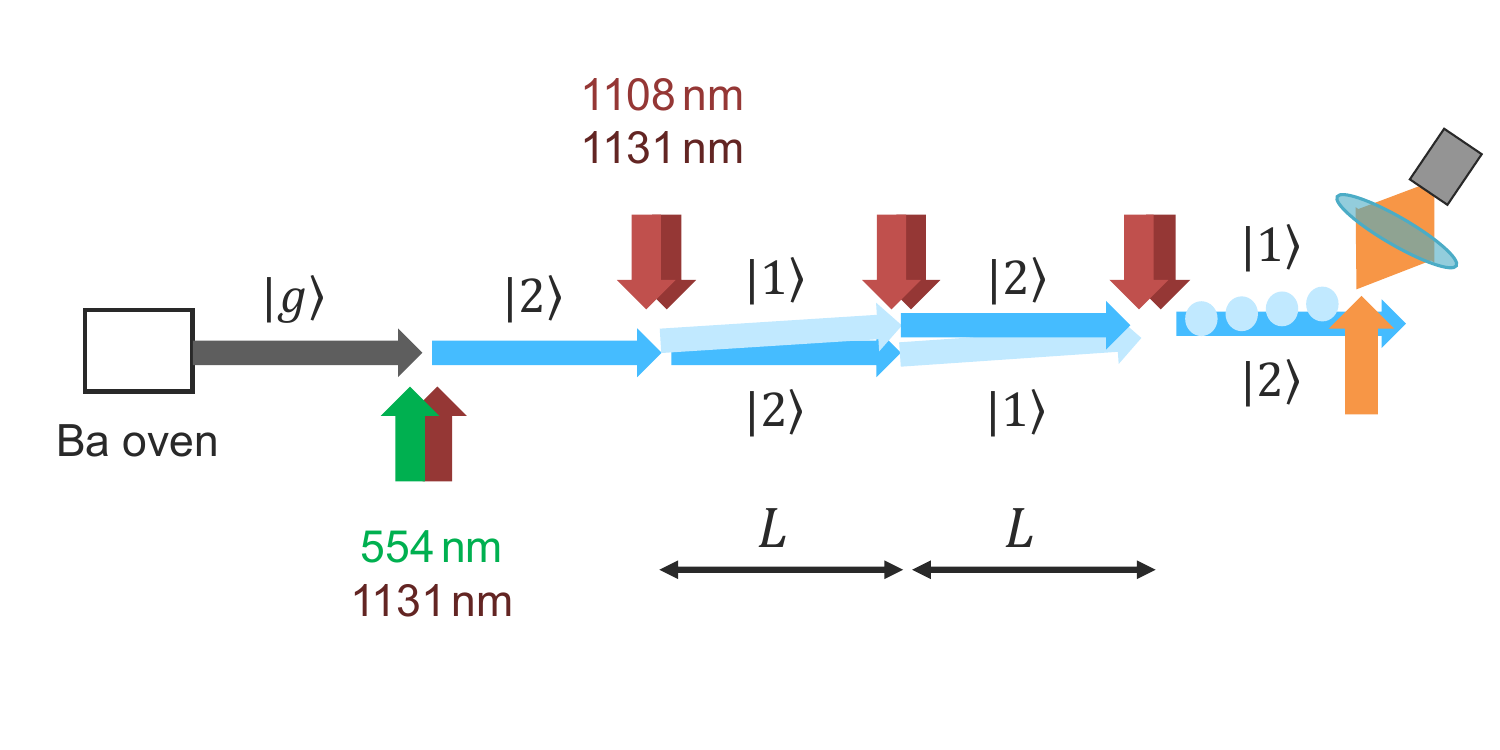}
\caption{
Experimental configuration of the co-propagating Mach--Zehnder atom interferometer based on metastable Ba atoms.
Three Raman interaction regions are formed by co-propagating 1108- and 1131-nm laser beams.
The separation between adjacent Raman beams is denoted by $L$, which is varied throughout the experiment.}
    \label{fig_experimental_MZ_config}
\end{figure}
\begin{figure}[htb]
    \centering
    \includegraphics[scale=0.42]{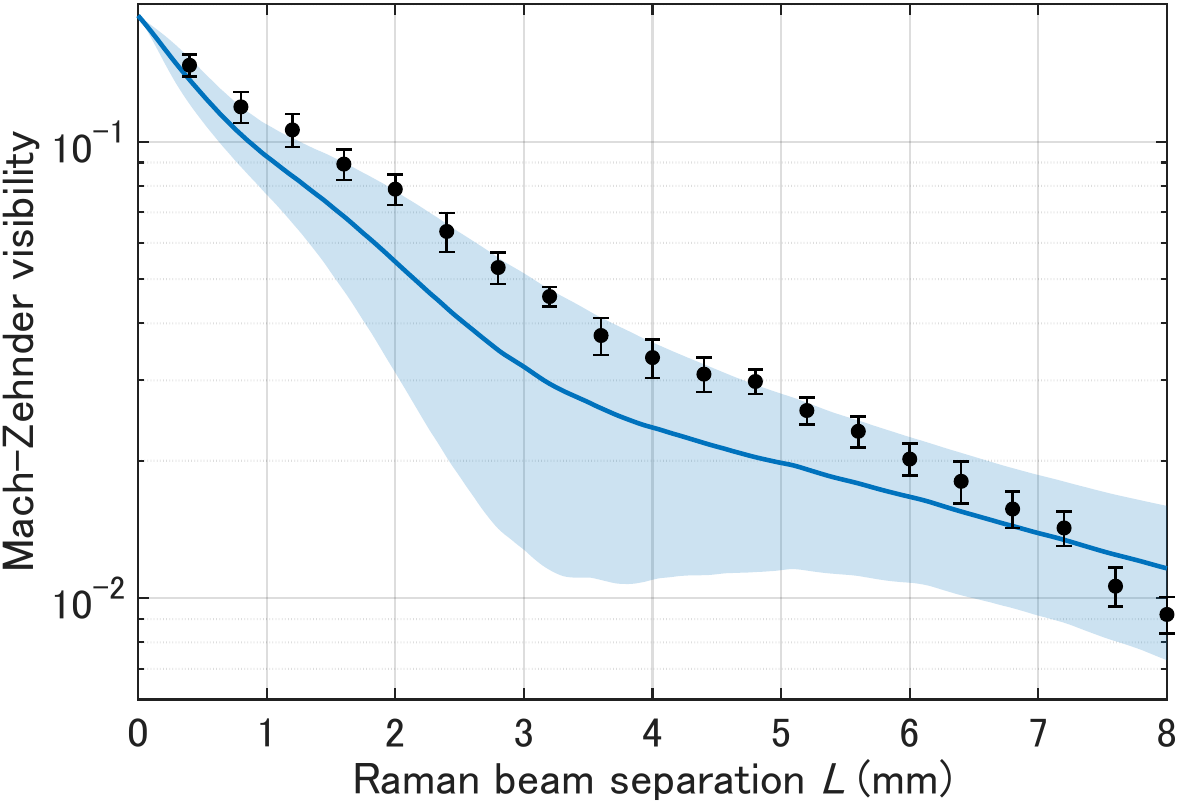}
\caption{
Mach--Zehnder visibility as a function of the Raman beam separation
$L$. The filled circles show the experimental data, while the blue
solid line shows the numerical simulation including the atomic velocity
distribution, the spatial intensity profiles of the Raman beams,
imperfect Raman pulse areas, spontaneous decay to the $|g\rangle$
state, and the measured three-point correlation of the relative phase
between the two Raman lasers.
The shaded region represents the
$\pm1\sigma$ uncertainty obtained from ten repeated measurements of
the three-point phase correlation.
}
    \label{fig_interference}
\end{figure}

The experimental configuration is shown in Fig.~\ref{fig_experimental_MZ_config}. The interferometer output was read out from the population in the $|1\rangle$ state using fluorescence on the $|1\rangle$--$|e^\prime\rangle$ transition at \SI{602}{nm}. Efficient fluorescence detection requires appropriate control of the magnetic field and probe polarization to avoid coherent population trapping, as described in Appendix~\ref{app:fluorescence_detection}.
The separation between adjacent Raman beams is denoted by $L$,
which was varied from $\SI{400}{\micro m}$ to \SI{8}{mm}.
The optical powers of the 1108-nm beams in the three Raman interaction
regions were 120, 140, and \SI{100}{mW}, respectively, while
the corresponding powers of the 1131-nm beams were
12, 13, and \SI{11}{mW}.
The $1/e^2$ beam radii in the three Raman interaction regions were
75, 80, and $\SI{85}{\micro m}$ along the atomic-beam direction and
2.9, 2.3, and \SI{2.8}{mm} along the direction perpendicular to both
the atomic and Raman beams, in the order of the first, second,
and third interaction regions.
The Raman transition was driven with a single-photon detuning of
$\Delta/2\pi=$\SI{-20}{MHz} from the intermediate state $|e\rangle$.
The $|1\rangle$--$|e\rangle$ transition is weak, with a transition rate of only
$2\pi\times$\SI{490}{Hz}, and the available 1108-nm laser power was
insufficient to realize the ideal $\pi/2$--$\pi$--$\pi/2$ pulse
sequence.
We therefore employed a relatively small one-photon detuning to obtain
sufficient Raman coupling.
For atoms with the most probable longitudinal velocity of \SI{360}{m/s},
as determined from the time-of-flight measurement described in Appendix~\ref{app:velocity_distribution},
the calculated loss-free Raman pulse areas for trajectories passing
through the centers of the Raman beams were $0.50\pi$, $0.73\pi$, 
and $0.48\pi$ 
in the three interaction regions, respectively.
The corresponding spontaneous-emission losses to the
$|g\rangle$ state were approximately \SI{31}{\percent},
\SI{34}{\percent}, and \SI{29}{\percent}.
In the experiment, a two-photon detuning of
$\delta/2\pi=\SI{140}{\hertz}$ was applied to the second Raman
interaction.
Under this condition, the Mach--Zehnder interference signal oscillates
at $2\delta$, corresponding to \SI{280}{\hertz}.
Residual Ramsey interference arising from the nonideal Raman pulse
areas appears at $\delta$ and is therefore spectrally separated from
the Mach--Zehnder signal.
Figure~\ref{fig_interference} shows the Mach--Zehnder visibility
as a function of the Raman beam separation $L$.
The experimental visibility was determined from the amplitude of the
280-Hz component relative to the DC fluorescence signal.
The filled circles represent the mean of ten repeated FFT measurements,
with error bars indicating one standard deviation (1$\sigma$),
while the solid line shows the numerical simulation.

To quantitatively model the dependence of the Mach--Zehnder
interference visibility on the Raman beam separation $L$, we performed
numerical simulations including the longitudinal velocity distribution
of the atomic beam, spontaneous-emission loss during the Raman
interactions, the spatial intensity profiles of the Raman beams, and
the relative phase fluctuations between the two Raman lasers.
The linewidths of the 1108- and 1131-nm Raman lasers were independently
measured using a delayed self-heterodyne method with a
$\SI{10}{\kilo m}$-long optical fiber, yielding linewidths of
$\SI{3}{\kilo\hertz}$ and $\SI{10}{\kilo\hertz}$, respectively.

If the laser frequency noise were purely white, the three-point phase
correlation relevant to the Mach--Zehnder sequence could be analytically
determined from the measured laser linewidths. In practice, however,
laser frequency noise can contain non-white components, for which the
linewidth alone does not fully characterize the temporal phase
correlations relevant to the Mach--Zehnder sequence.
We therefore directly measured the three-point phase correlation of
the relative phase between the two Raman lasers.
For this measurement, the beat note between each Raman laser and the
common optical frequency comb was generated separately, and the two
beat signals were mixed using a double-balanced mixer to obtain their
difference-frequency signal. In this configuration, phase fluctuations
of the common optical frequency comb are canceled as a common-mode
contribution, so that the resulting signal directly reflects the
relative phase fluctuations between the 1108- and 1131-nm Raman
lasers. The difference-frequency signal was measured using a frequency
counter, and the three-point phase correlation corresponding to the
Mach--Zehnder sequence was evaluated as a function of the time interval
between successive Raman interactions.
The measurement was repeated ten times, and the mean three-point phase
correlation was used in the numerical simulation. The standard
deviation of the ten measurements was propagated through the simulation
to evaluate the uncertainty arising from the measured relative phase
fluctuations. Together with the experimentally determined longitudinal
velocity distribution and independently measured experimental
parameters, these measurements were used in the simulation without
any adjustable parameters.

As shown in Fig.~\ref{fig_interference}, the numerical simulation
quantitatively reproduces the overall magnitude of the visibility and its
dependence on $L$. At the shortest Raman-beam separations, Mach--Zehnder
visibilities exceeding \SI{10}{\percent} were observed despite spontaneous-emission
loss to the $|g\rangle$ state and the nonideal Raman pulse areas.
The subsequent decrease in visibility with increasing $L$ is predominantly
accounted for by the measured phase fluctuations of the Raman lasers.
A detailed characterization of the laser phase noise and the
three-point-correlation measurement method is beyond the scope of the
present work and will be reported separately.

\section{conclusion\label{section-conclusion}}
In conclusion, we have proposed and experimentally demonstrated a
Mach--Zehnder atom interferometer based on coherent Raman transitions
between two metastable fine-structure states of Ba atoms.
Mach--Zehnder visibilities exceeding \SI{10}{\percent} were observed at short
Raman-beam separations, with interference remaining observable over
separations of several millimeters. The overall magnitude and dependence
of the visibility on the Raman beam separation are quantitatively
reproduced by our numerical simulation without adjustable parameters,
incorporating the experimentally measured three-point correlation of the
relative phase between the two Raman lasers. This agreement supports the
interferometric mechanism demonstrated in this work and indicates that
the reduction in visibility at larger Raman beam separations is
predominantly limited by the relative phase fluctuations between the
Raman lasers.

For practical inertial sensing, the Raman beam separation must be
extended from the present millimeter scale to approximately
$\SI{100}{\milli\meter}$.
Maintaining high interference visibility over such an extended
interaction region will require both stronger Raman laser fields and
improved relative phase coherence.
Higher Raman laser power would allow operation at a larger one-photon
detuning while maintaining sufficient two-photon Raman coupling,
thereby suppressing spontaneous-emission loss.
In addition, phase locking the 1108- and 1131-nm Raman lasers to the
common optical frequency comb is expected to suppress their relative
phase fluctuations and enable improved visibility to be maintained at
larger Raman beam separations.
Although the present proof-of-principle experiment employed
co-propagating Raman beams, practical inertial sensors can be realized
using counter-propagating Raman beams generated from a common optical
path by retroreflection, thereby preserving the relative optical phase
while providing the large effective wave vector required for high
inertial sensitivity.
Experimental studies toward a counter-propagating implementation for
inertial sensing, including direct evaluation of the predicted magnetic
robustness, are currently in progress.
These developments are expected to establish fine-structure Raman
interferometry as a new architecture for atom interferometers,
combining the versatility of Raman-based operation with intrinsically
low magnetic sensitivity for practical quantum inertial sensing.

\section*{Acknowledgments}
We are grateful to K. Tanaka for invaluable experimental support and to K. Aikawa for valuable assistance with the atomic-structure calculations. This work was supported by JST, Japan, under Grant Number JPMJPF2015.

\section*{Data availability}
The data that support the findings of this article are openly available~\cite{dataset}.

\appendix

\section{State-selective fluorescence detection of metastable Ba atoms}
\label{app:fluorescence_detection}

Coherent population trapping among the three Zeeman sublevels ($m_J=-1,0,+1$) of $|1\rangle$ and the single sublevel ($m_J=0$) of $|e^\prime\rangle$ state leads to the formation of dark states, hindering efficient fluorescence detection. Following Ref.~\cite{Hemingway2020}, we apply a magnetic field and mix the probe polarizations to destabilize the dark states.

\begin{figure}[htb]
    \centering
    \includegraphics[scale=0.28]{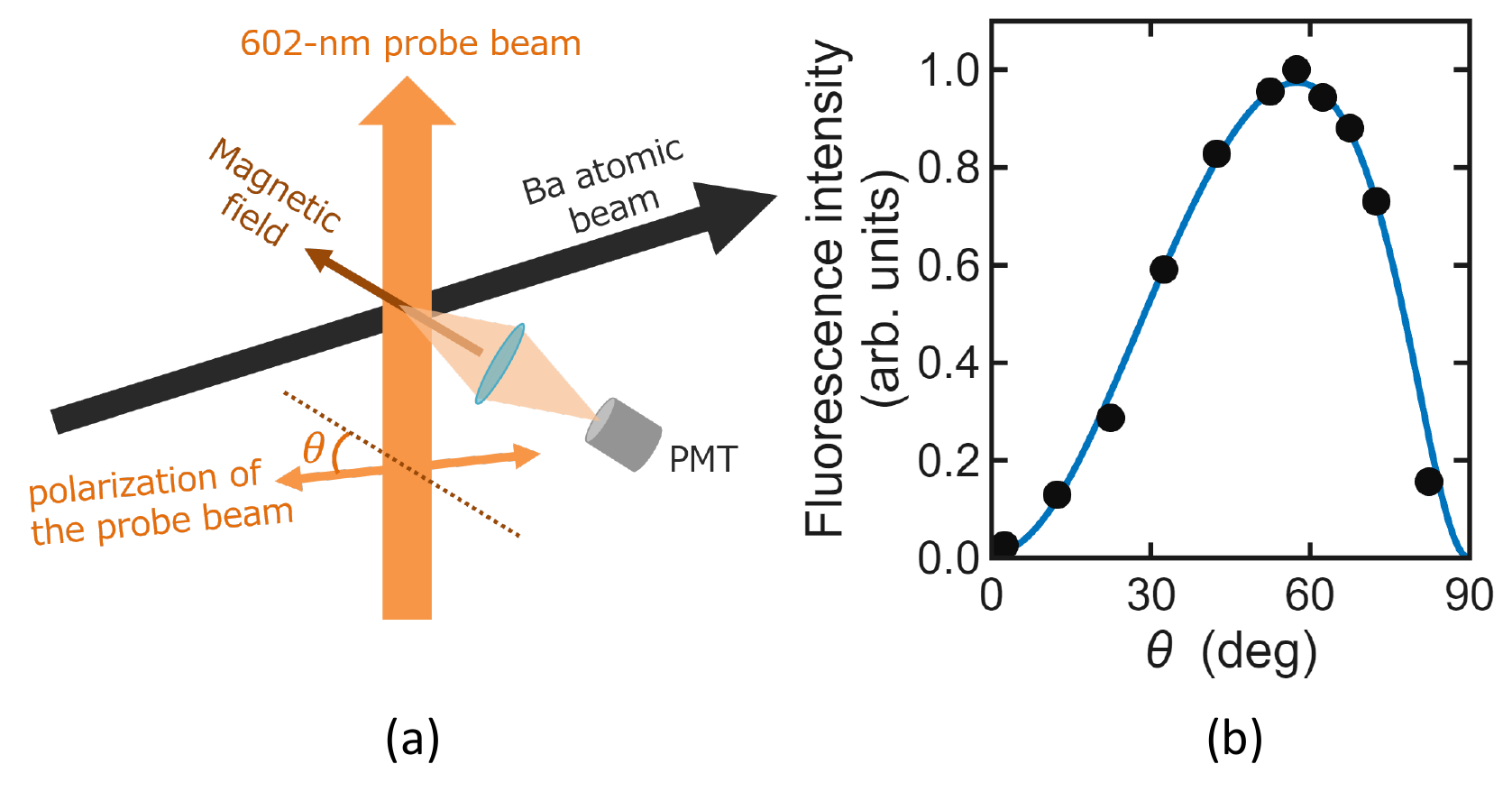}
\caption{
State-selective fluorescence detection of metastable Ba atoms.
(a) Experimental setup. The angle $\theta$ is defined between the magnetic field and the linear polarization of the 602-nm probe beam. Fluorescence emitted along the magnetic-field direction is detected with a photomultiplier tube (PMT).
(b) Fluorescence intensity as a function of $\theta$. The black filled circles show the measured fluorescence signal, normalized to its maximum value, and the blue solid curve shows the optical Bloch equation simulation.
}
    \label{fig_602detection}
\end{figure}
Figure~\ref{fig_602detection}(a) shows the experimental setup.
A 26-mW, 602-nm probe beam with a $1/e^2$ radius of \SI{12}{\milli \meter} intersects the metastable Ba atomic beam at right angles.
A magnetic field of \SI{160}{\micro\tesla} is applied perpendicular to both the atomic beam and the probe beam, such that the three directions are mutually orthogonal.
By rotating the probe polarization, the relative contributions of the $\pi$- and $\sigma^{\pm}$-polarized components are varied.
Figure~\ref{fig_602detection}(b) shows the measured fluorescence intensity as a function of $\theta$.
The fluorescence intensity reaches its maximum at $\theta \approx \SI{57}{\degree}$, in good agreement with the optical Bloch equation simulation.

\begin{figure}[htb]
    \centering
    \includegraphics[scale=0.64]{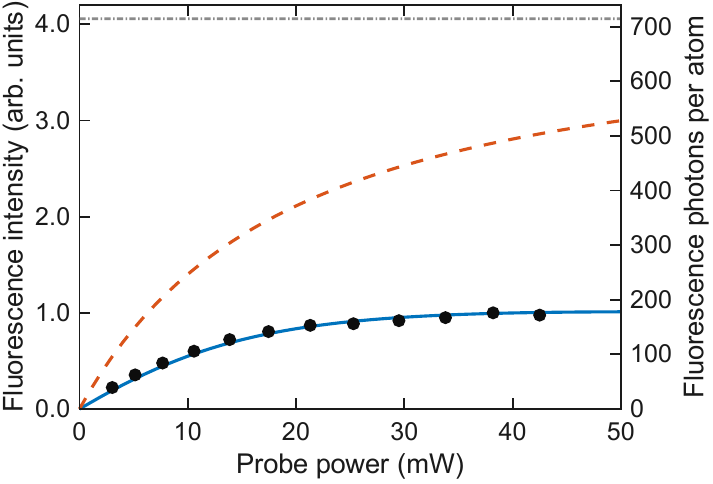}
\caption{
Dependence of the 602-nm fluorescence signal on the probe laser power.
The black filled circles show the measured fluorescence signal.
The blue solid curve shows the optical Bloch equation simulation for
the actual four-level system, with its overall amplitude scaled to fit
the experimental data.
The orange dashed curve shows the calculated fluorescence yield for an
ideal two-level system, and the gray dash-dotted line indicates the
leakage-limited maximum of 710 fluorescence photons per atom.
The left and right vertical axes represent the measured fluorescence
signal and the simulated fluorescence photons per atom, respectively.
}
    \label{fig_602power}
\end{figure}
To assess whether a sufficient number of fluorescence photons can be obtained to enable atom-shot-noise-limited detection, we measured the dependence of the 602-nm fluorescence signal on the probe laser power. Figure~\ref{fig_602power} shows the measured fluorescence signal under the optimized polarization condition described above. The overall amplitude of the simulated fluorescence signal is scaled to fit the experimental data, while its dependence on the probe power is determined by the simulation. The simulated power dependence is in good agreement with the experimental results over the entire range of probe powers, supporting the validity of the optical Bloch equation model for describing the fluorescence dynamics of the four-level system.

Using this model, we estimate the number of fluorescence photons per atom under the experimental conditions, as shown on the right vertical axis. For comparison, the orange dashed curve shows the calculated fluorescence yield for an ideal two-level system. Although the fluorescence yield of the actual four-level system is lower than that of the ideal two-level system, the simulation predicts that several hundred fluorescence photons per atom can still be obtained under the present experimental conditions. The gray dash-dotted line indicates the theoretical upper limit of 710 fluorescence photons per atom, determined by the leakage branching ratio of $1.4\times10^{-3}$ from $|e^\prime\rangle$. Relativistic CI+MBPT calculations performed using the package developed by Kozlov \textit{et al.}~\cite{Kozlov2015CIMBPT} predict that the dominant leakage channel is spontaneous decay to the nearby ${}^{3}P_{1}\,(5d^{2})$ state.

Despite the multilevel structure and associated coherent population trapping, the predicted fluorescence yield of several hundred photons per atom indicates that fluorescence photon shot noise need not prevent atom-shot-noise-limited detection, provided that sufficient photon collection and detection efficiencies are achieved.

\section{Longitudinal velocity distribution of the Ba atomic beam}
\label{app:velocity_distribution}

\begin{figure}[htb]
    \centering
    \includegraphics[scale=0.34]{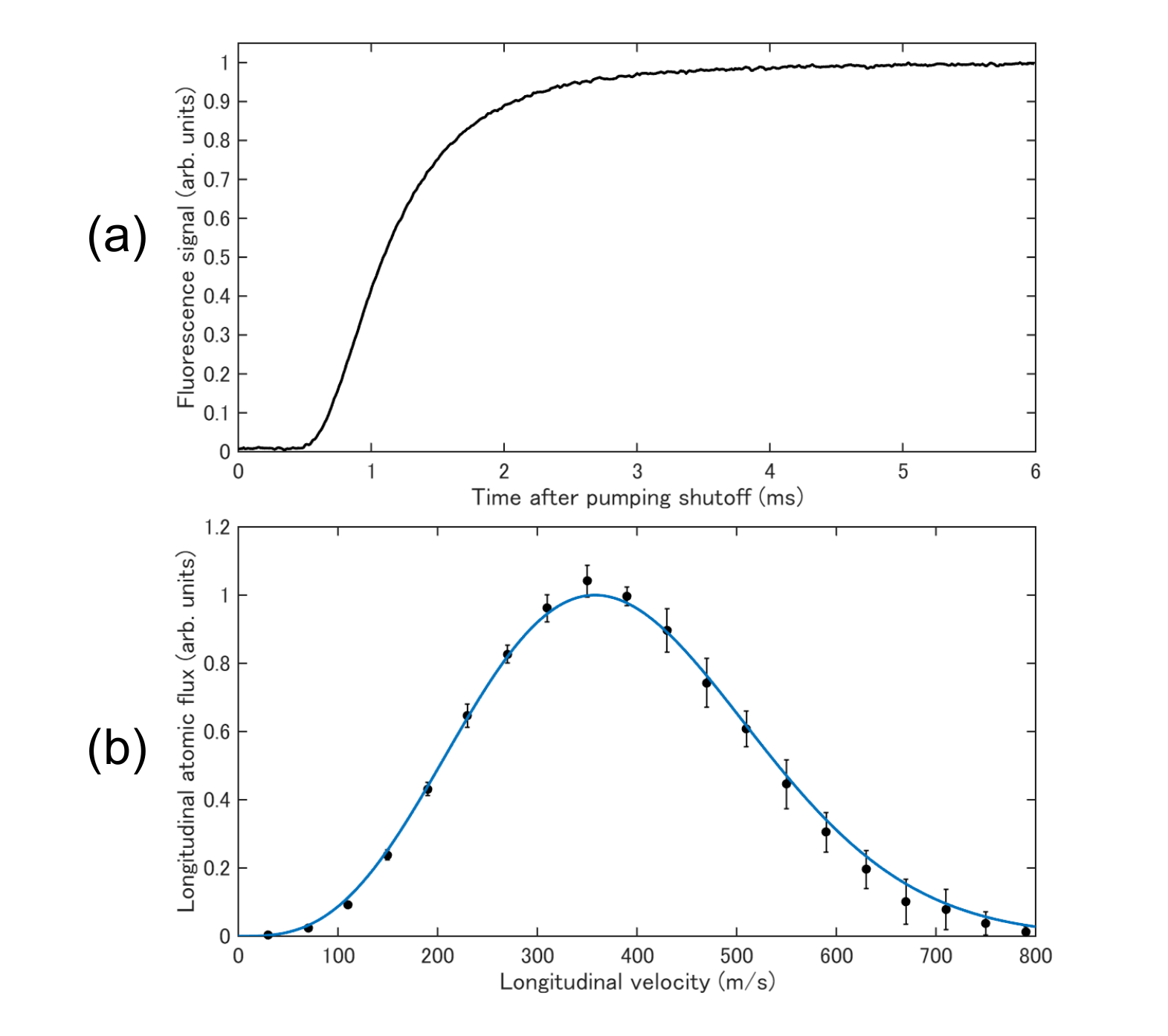}
    \caption{
    Time-of-flight characterization of the longitudinal velocity
    distribution of the Ba atomic beam.
    (a) Time evolution of the 554-nm fluorescence signal after the
    optical-pumping lasers were switched off.
    (b) Longitudinal velocity distribution obtained from the time
    derivative of the signal in (a).
    The points show the experimental data, and the solid curve is a fit
    to the effusive-oven velocity distribution, proportional to
    $v_{l}^3\exp[-mv_{l}^2/(2k_{\mathrm B}T)]$, where $v_{l}$ is the longitudinal
    atomic velocity, $m$ is the Ba atomic mass, $k_{\mathrm B}$ is the
    Boltzmann constant, and $T$ is the oven temperature.
    }
    \label{fig_longitudinal_velocity}
\end{figure}

The longitudinal velocity distribution of the atomic beam was
characterized by a time-of-flight measurement
\cite{Park2012,hosoya2023}.
The 554- and 1131-nm lasers used for optical pumping into the metastable
state were abruptly switched off, and the subsequent rise of the 554-nm
fluorescence signal from ground-state atoms was recorded at a detection
region located \SI{360}{mm} downstream from the optical-pumping beams, as
shown in Fig.~\ref{fig_longitudinal_velocity}(a).
The velocity distribution obtained from the time derivative of the
fluorescence signal is shown in Fig.~\ref{fig_longitudinal_velocity}(b).
The measured distribution is well described by the effusive-oven model,
yielding a most probable longitudinal velocity of approximately
\SI{360}{m/s} (oven temperature of \SI{440}{\celsius}).

\clearpage

\bibliography{references}

\end{document}